\documentclass[aps,twocolumn,superscriptaddress,longbibliography,nofootinbib]{revtex4-1}
\usepackage{graphicx}
\usepackage{bm,amsmath}
\usepackage{dcolumn}
\usepackage{amsfonts,amssymb}
\usepackage{hyperref}
\usepackage{enumitem}
\usepackage{slashed}

\newcommand{\tr}{\mathrm{tr}\,}
\newcommand{\Tr}{\mathrm{Tr}\,}
\newcommand{\CC}{{\cal C}}

\newcommand{\eqn}[2] {\begin{equation} \label{#1} #2 \end{equation}}

\def\C{\mathbb{C}}
\def\P{\mathbb{P}}
\def\Z{\mathbb{Z}}
\def\R{\mathbb{R}}
\def\foot{\footnote}

\newcommand{\be}{\begin{equation}}
\newcommand{\ee}{\end{equation}}

\begin{document}
\title{
Walls, Anomalies, and (De)Confinement in Quantum Anti-Ferromagnets
}
\author{Zohar Komargodski}
\email{zohar.komargodski@weizmann.ac.il}
\affiliation{ Department of Particle Physics and Astrophysics, Weizmann Institute of Science, Israel}
\affiliation{Simons Center for Geometry and Physics, Stony Brook University, Stony Brook, NY}

\author{Tin Sulejmanpasic} 
\email{tin.sulejmanpasic@gmail.com}
\affiliation{Philippe Meyer Institute, Physics Department, \'Ecole Normale Sup\'erieure, PSL Research University, 24 rue Lhomond, F-75231 Paris Cedex 05, France}
\author{Mithat \"Unsal} 
\email{unsal.mithat@gmail.com}
\affiliation{Department of Physics, North Carolina State University, Raleigh, North Carolina 27695, USA}

\date{\today}

\vskip 2 cm

\begin{abstract}

We consider the Abelian-Higgs model in 2+1 dimensions with instanton-monopole defects. This model is closely related to the phases of quantum anti-ferromagnets. In the presence of $\mathbb{Z}_2$ preserving monopole operators, there are two confining ground states in the monopole phase, corresponding to the Valence Bond Solid (VBS) phase of quantum magnets. We show that the domain-wall carries a 't Hooft anomaly in this case. The anomaly can be saturated by, e.g., charge-conjugation breaking on the wall or by the domain wall theory becoming gapless (a gapless model that saturates the anomaly is $SU(2)_1$ WZW). Either way the fundamental scalar particles (i.e. spinons) which are confined in the bulk are deconfined on the domain-wall. This $\Z_2$ phase can be realized either with spin-1/2 on a rectangular lattice, or spin-1 on a square lattice. In both cases the domain wall contains spin-1/2 particles (which are absent in the bulk). We discuss the possible relation to recent lattice simulations of domain walls in VBS. We further generalize the discussion to Abrikosov-Nielsen-Olsen (ANO) vortices in a dual superconductor of the Abelian-Higgs model in 3+1 dimensions, and to the easy-plane limit of anti-ferromagnets. In the latter case the wall can undergo a variant of the BKT transition (consistent with the anomalies) while the bulk is still gapped. The same is true for the easy-axis limit of anti-ferromagnets. We also touch upon some analogies to Yang-Mills theory.

\end{abstract}
\maketitle

\section{Introduction}

't Hooft anomaly matching \cite{tHooft:1979rat} is a powerful method which imposes strong constrains on the infrared (IR) physics. The method relies on the observation that certain global symmetries resist promotion to gauge symmetries. In other words, a gauge formulation of the theory is not possible without an introduction of a non-local counter-term depending on the background gauge fields  which cancels the gauge non-invariance of the original path integral. The gauge variation of the counter-term, being a c-number, can be performed at all energy scales, and, in particular, at low energies. Therefore, the effective theory at all scales has to reproduce the same transformations under gauge variations of the background gauge fields.\foot{A point which we do not discuss in detail but will be implicit throughout is that some of the symmetries that have 't Hooft anomalies typically appear as discrete spatial symmetries in the underlying spin system on the lattice (see e.g. \cite{Read:1990zza}). This is not surprising, since if the symmetries were manifest, on-site, symmetries, we could gauge them as usual by adding gauge fields on the links.}

The traditional 't Hooft anomalies involve continuous symmetries and continuous anomalies. However, recently it became clear that 't Hooft anomalies are much more generic, and may include discrete symmetries as well. For some recent work, see, for example, \cite{Kapustin:2014lwa,Gaiotto:2014kfa,Benini:2017dus, Gaiotto:2017yup, Wang:2017txt, Komargodski:2017dmc, Wang:2017loc,Tanizaki:2017bam} and references therein (see also \cite{Wang:2014pma,Wen:2014zga} for a related discussion in the context of condensed matter physics). Such 't Hooft anomalies transcend theories with fermions and are not restricted to even dimensions. 

Loosely speaking, continuous 't Hooft anomalies are matched with either Nambu-Goldstone bosons or with (possibly free) conformal field theories containing fermions~\cite{Frishman:1980dq}. 
Likewise, discrete 't Hooft anomalies can be matched by either breaking some symmetry (which leads to multiple degenerate ground states) or by having a nontrivial low-energy theory in a symmetric vacuum. In condensed matter the anomaly is perhaps best known as being responsible for the edge modes of the 2+1D $U(1)$ Chern-Simons (CS) theory, where gauge invariance of the CS Lagrangian requires the existence of chiral edge modes to cancel the anomaly inflow~\cite{Callan:1984sa, Faddeev:1985iz} from the bulk. 

On the other hand when ordinary discrete (i.e. 0-form\footnote{See \cite{Gaiotto:2014kfa} for a discussion of $p$-form symmetries}) symmetries are spontaneously broken, the system supports domain-wall excitations. These are topologically stable co-dimension 1 objects which interpolate between the different vacua. It was recently shown that several semi-classical regimes (in Yang-Mills theories and anti-ferromagnets) with spontaneously broken discrete symmetries support excitations which are confined in the bulk, and hence absent from the bulk spectrum \cite{Anber:2015kea,Sulejmanpasic:2016uwq}, a behavior conjectured a long time ago to hold in ${\cal N}=1$ Super Yang-Mills theory \cite{Witten:1997ep}. This deconfinement on the wall was also argued and verified numerically in the non-semiclassical regime of spin-$1/2$ antiferromagnets \cite{Sulejmanpasic:2016uwq}, giving evidence that the phenomenon is robust. Since the domain wall can be thought of as the boundary separating two bulk states, the situation is reminiscent of the existence of edge modes in topological insulators~\cite{Diamantini:2016mrb}. This is not an accident, and the existence of 't Hooft anomalies in the global (discrete and continuous) symmetries of the relevant systems is crucial for the appearance of the domain-wall modes, otherwise absent in the bulk.

This mechanism is briefly described as follows. The domain walls are the analogs of Nambu-Goldstone bosons for discrete symmetries. On a compact manifold such a domain-wall can be thought of as arising because the boundary conditions along one of the directions of space are twisted by a symmetry transformation or, equivalently, by turning on the background gauge fields for the discrete symmetry. The natural question which then arises is what is the theory on the domain-wall and what are its properties. Oftentimes, the original anomaly in the bulk forces the domain wall to be nontrivial in order for the anomaly to be matched. This is therefore a discrete version of the anomaly inflow mechanism. 

While we mostly focus on the study of the Abelian Higgs model in 2+1 dimensions with monopole operators added to the Lagrangian, we also give an overview of similar phenomenon in the 3+1D Yang-Mills theories in the appendix as an analogy. The bulk of the work is dedicated to the so-called Valence Bond Solid (VBS) phase of 2+1D anti-ferromagnets, which is descried by the Abelian-Higgs model\footnote{See also~\cite{Haldane:1982rj,Baskaran:1987my,haldane1988,Haldane:1983ru,Dzyaloshinsky:1987pq,WIEGMANN1988103,Wiegmann:1988qn,Fradkin:1989hj} for related works.}~\cite{senthil2004deconfined}  in the regime where the charged particles are confined by the Polyakov mechanism. If only even monopoles are present, there are two ground states. There exists, therefore, a domain wall interpolating between these vacua. Due to 't Hooft anomalies in the bulk, the domain wall theory also carries 't Hooft anomalies. For example, in the most familiar case with two charged particles with a $\mathbb{C}\mathbb{P}^1$ N\'eel phase, the domain wall in the VBS phase carries a $\pi i\int w_3(O(3))$ anomaly, where $w_3$ is the 3rd Stiefel-Whitney class. We find that the domain wall itself has two ground states, which saturate the above anomaly. The charged particles, which are confined in the bulk, fractionalize and become deconfined on the wall and can be viewed as a domain wall within a domain wall.
We further suggest that the domain wall may become critical away from the semi-classical regime (while the bulk is still gapped) and argue that the anomaly can be matched by the $SU(2)_1$ WZW model. This conjecture is consistent with the recent Monte Carlo simulations \cite{Sulejmanpasic:2016uwq}, where the domain wall theory of the $\Z_2$ VBS appears to be critical and consistent with the spin-1/2 chain, which is in the same universality as $SU(2)_1$ WZW model.

We similarly analyze the easy-plane and easy-axis modifications of the model, showing that again the domain wall may undergo an interesting phase transition while the bulk is still gapped. For the easy-plane model, in the semi-classical regime the wall has two ground states and deconfined spinons. The domain wall may become massless as we move into the strongly coupled regime undergoing an interesting modification of the BKT transition. In the easy-axis case, the domain wall exists when the bulk is really in the N\'eel phase, and the domain wall is massless already semi-classically. It may become massive in the quantum regime, again undergoing a modified BKT-like transition. We also discuss some analogies between the physics of these anti-ferromagnets, Yang-Mills theory, and  the dynamics of ANO strings in 3+1 dimensions.

\section{The 2+1D Abelian-Higgs Model}\label{sec:abelian-higgs}

\subsection{Preliminaries}

The Lagrangian of this model is given by
\begin{multline}\label{LagAH}
{\cal L}=-{\frac{1}{4e^2} }|da|^2+\sum_i |D_a\phi_i|^2\\+m^2\sum_i|\phi_i|^2+\lambda\left(\sum_i |\phi_i|^2\right)^2~.
\end{multline}
 The symmetries of this model are 
\eqn{symms}{\CC \ltimes (PSU(N_f)\times U(1)_T)~.}
$\CC$ stands for charge conjugation, $PSU(N_f)$ 
is the group that rotates the fields $\phi_i$ (which we call spinons) and the $U(1)_T$, where $T$ stands for \emph{topological}, reflects the topologically conserved current $j=\star F_a$ which is conserved by the Bianchi identity in the theory without monopoles added to the Lagrangian. The conserved charge of $U(1)_T$ is the topological charge $Q = \int_2 da/2\pi$, or the skyrmion number. Note that the flavor group is $PSU(N_f)=SU(N_f)/\mathbb Z_{N_f}$ and not $SU(N_f)$ because the center elements of $SU(N_f)$ are $U(1)$ gauge transformations.

Motivated by observations made in \cite{Sulejmanpasic:2016uwq}, we will be interested in the model~\eqref{LagAH} where only $\Z_2\subset U(1)_T$ is a symmetry. From the point of view of~\eqref{LagAH} this means that we imagine adding monopole operators with  even charges under $U(1)_T$.
The symmetry group now is therefore
\eqn{symmsi}{(\CC\ltimes PSU(N_f))\times \Z_2~.}
$\Z_2$ now appears in a direct product because it commutes with charge conjugation.

The model with even monopoles, and with the symmetry group~\eqref{symmsi} has a 't Hooft anomaly. Below we review the argument~\cite{Komargodski:2017dmc} that the above symmetry cannot be gauged in a consistent manner; one is forced to add non-local counter terms to restore gauge invariance. 

Let us first introduce some notation. We will think about $PSU(N_f)$ gauge fields as follows. One element will be  the standard gauge fields of $SU(N_f)$ which we denote as $a'$, while the other element will be a 2-form $\Z_{N_f}$-valued gauge field $B$ associated with the gauging of the center symmetry of $SU(N_f)$ to obtain $PSU(N_f)$ (see e.g. \cite{Gaiotto:2014kfa,Kapustin:2014gua} for further details about this approach to $PSU$ gauge fields). We normalize the field $B$ as $e^{i\oint B}=e^{i\frac{2\pi k}{N_f}},\;k\in \mathbb Z$, where the integral is over an arbitrary 2-cycle. Note that the holonomies $e^{ip\oint B}$ are well defined only if $p\in \mathbb Z$ and where $p$ and $p+N_f$ are equivalent. This is because $B$  is a gauge field, so that shifts $B\rightarrow B+d\lambda$ where $\lambda$ can be thought of as a properly normalized $U(1)$ gauge field, are gauge transformations, which then induce $\oint_2 B\rightarrow \oint_2 B+2\pi k, k\in \Z$. 

Finally, the integral over $B$ with a boundary is only well defined if there is an object on the boundary which can cancel the gauge transformation of $B$. For example if we integrate $B$ over a disc $D$ we get that the gauge variation of $\int_D B$ is $\int_{\partial D}\lambda$. The object which cancels this term is the fundamental Wilson loop of the $SU(N_f)$ gauge fields $a'$ (see e.g. \cite{Gaiotto:2014kfa}). So if an integral over the $B$-field appears, it means that at the boundary of the integration domain we have $SU(N_f)$ charges which are not $PSU(N_f)$ charges. In other words, at the boundary we have a projective representation of $PSU(N_f)$.

In addition we will further denote the $\Z_2$-valued 1-form gauge field by $A$ and normalized such that $e^{i\oint A}=\pm1$, and similarly for the charge-conjugation $\CC$-symmetry we introduce the 1-form gauge field $C$.

\subsection{Semiclassical analysis of the VBS phase}

 Let us now consider a VBS (Polyakov) phase of \eqref{LagAH} with $m\gg e^2$, i.e in the semi-classical limit. Such a theory is a pure gauge theory described approximately by the first term of $\eqref{LagAH}$. However, we assume in addition that there are even monopole operators in the system, coming from the UV theory. We denote the space-time density of these monopoles by $\rho_m$ (without loss of generality, we choose $\rho_m$ to be positive). This theory can be dualized to a single compact scalar:
\be\label{eq:sigma_lag}
\mathcal L_{eff}=\frac{e^2}{2(2\pi)^2}(\partial_{i}\varphi)^2-\rho_m\cos(2\varphi)~,
\ee
where $\varphi\sim \varphi+2\pi$ is a compact scalar -- the dual photon -- and where we assumed that the charge of the monopoles in the Lagrangian is $\pm 2$, preserving the $\Z_{2}\subset U_T(1)$.\footnote{In quantum anti-ferromagnets the discrete topological symmetry is identified with some subgroup of the discrete lattice symmetries. See e.g. \cite{Read:1990zza}. In order to realize our model on the lattice in spin-$1/2$ systems, all one needs to do is to consider rectangular lattices, where 90 degree rotations are not a symmetry. Many of the statements below can be understood also on the lattice. Alternatively we also argue that the VBS phase of the spin-$1$ system has similar features, as we discuss below. We thank C.~Wang for a discussion on this.} (We could add also higher monopoles with even charge. We will make some comments about this scenario below.) The above description is valid as long as $\rho_m/e^6\ll 1$, i.e. that the typical separation between monopoles is much smaller than the length scale at which $\varphi$ develops a long-range order. In that case the monopole contribution pins the $\varphi$ dual photon down at either $\varphi=0,\pi \mod 2\pi$, yielding two distinct vacua. The symmetries $\CC$ and $\Z_{2}$-topological therefore act as
\be
\Z_{2}: \; \varphi\rightarrow \varphi+\pi\;,\qquad \CC: \; \varphi\rightarrow -\varphi\;.
\ee
The low-energy field $\varphi$ does not transform under the $SO(3)$ symmetry of the spinons (we focus here of $N_f=2$). The two vacua at $\varphi=0,\pi \mod 2\pi$ preserve charge conjugation but break the $\Z_2$ spontaneously. We will see that in spite of the fact that the $SO(3)$ quantum numbers are not visible in~\eqref{eq:sigma_lag}, by studying domain walls one rediscovers the $SO(3)$ quantum numbers. 

Indeed, since the $\Z_2$ symmetry is spontaneously broken, the effective theory allows for domain-walls between the two vacua. In fact, in the semi-classical regime where~\eqref{eq:sigma_lag} is valid, there are two distinct ground states for the domain-wall, denoted by $DW_\pm$, which are related by the $\CC$ symmetry,\footnote{These differ by whether $\varphi=0\mod 2\pi$ winds to $\varphi=\pi\mod 2\pi$ with a positive or negative gradient as the wall is crossed.} so that the wall spontaneously breaks the $\CC$ symmetry, but restores the $\Z_2$-topological symmetry.\footnote{$\cos(\varphi)$ is an order parameter for $\Z_2$ and
$\sin(\varphi)$ is an order parameter for $\CC$ symmetry. As the field profile interpolates from $\varphi=0$ to $\varphi=\pi$, it is clear that 
off the wall  $\Z_2 \times  \CC \rightarrow \CC$ and on the wall $\Z_2 \times \CC \rightarrow \Z_2$, i.e, either one or the other is broken, consistent with anomaly.} 

We could also consider more general potentials, involving a sum over even frequencies 
$V=-\sum_k\rho^k_m\cos(2k\varphi)$. Such potentials, unless they are fine tuned, have either 2 or 4 ground states. The case of 2 ground states always breaks $\Z_2$ and preserves $\CC$. The case of four ground states breaks both $\CC$ and $\Z_2$. It has different domain walls about which we will make a few comments below. For the time, we will discuss the scenario with two ground states, where $\Z_2$ is broken and $\CC$ is preserved. As we mentioned, semi-classically, it is obvious that $DW_\pm$ are degenerate and hence $\CC$ is broken on the wall even though it is unbroken in the bulk.

\begin{figure}[tbp] 
   \centering
   \includegraphics[width=.5\textwidth]{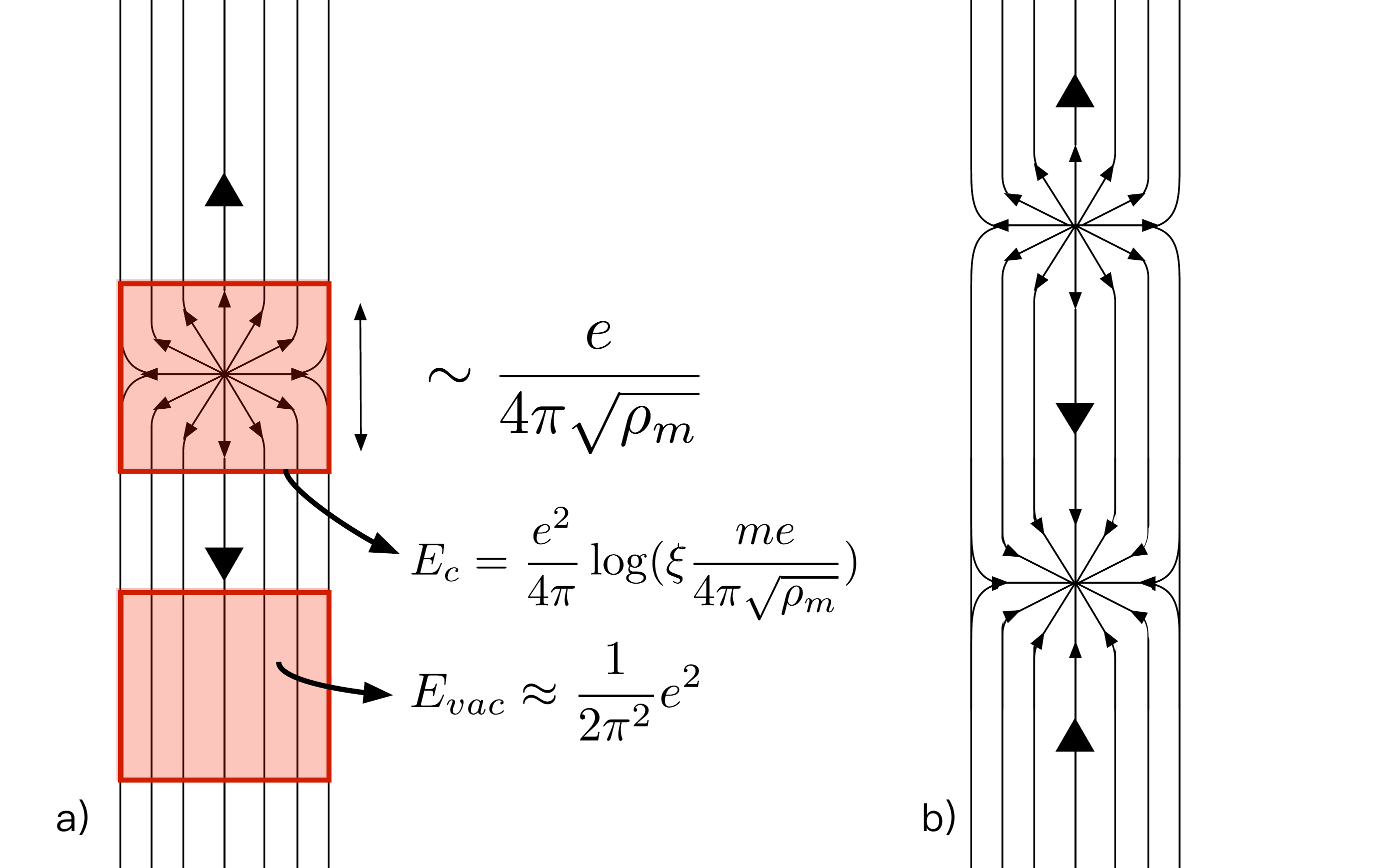} 
   \caption{The schematic depiction of domain-walls in $\Z_2$ broken VBS. The domain-walls restore the topological symmetry, but break charge conjugation because they carry electric flux (depicted by the arrows). The domain-walls inside the domain-walls are elementary excitations of spinons which do not exist in the bulk as they are confined. In $a)$ a single spinon on the domain-wall is depicted, as well as terms which enter \eqref{eq:meff}. In $b)$ a pair created from the vacuum of the domain-wall is shown, changing the vacuum in between them by the charge-conjugation (i.e. changing the flux on the domain wall).}
   \label{fig:DW}
\end{figure}

Let us study the semi-classical properties of $DW_\pm$ a little further. The domain-wall has tension given by $2e\sqrt{\rho_m}$. A domain-wall inside the domain-wall is an elementary scalar excitation carrying charge under the $U(1)$ gauge field, depicted in Fig. \ref{fig:DW}a). Such an object has a bare energy $E=m$, where $m$ is the mass of the scalar excitation. However it locally distorts the electric field lines $E_i\propto \epsilon_{0ij}\partial_j\varphi$. Approximately the presence of the $\phi$-excitation on the domain-wall has an energy difference from the ground state  (i.e. an effective mass) given by
\be\label{eq:meff}
m_{eff}^{DW}=m-\frac{e^2}{2\pi^2}+\frac{e^2}{4\pi}\log(\xi\frac{me}{4\pi \sqrt{\rho_m}})
\ee
where the second and third term are $E_c$ and $E_{vac}$ depicted in the shaded region of Fig. \ref{fig:DW} and $\xi$ is a number of order unity. We have not indicated here, but the coupling $e$ and the monopole density $\rho_m$ are expected to change with the mass $m$, which provides the UV cutoff of the effective theory. In the large $m$ limit the mass of the $\phi$ excitation is clearly positive. However as $m$ is decreased the effective mass on the domain-wall decreases. Two scenarios are possible:
\begin{itemize}
\item As $m$ is lowered, the effective mass $m_{eff}$ of $\phi$ on the domain-wall does not vanish before the bulk theory undergoes a transition.
\item The effective mass of $\phi$ on the domain-wall goes to zero while the bulk is still gapped.
\end{itemize}

If the first scenario is realized, the theory on the wall is always gapped and breaks the $\CC$-symmetry and, hence, supports stable domain-walls -- deconfined $\phi$ excitations -- which carry $SU(2)$ quantum numbers and are not present in the bulk. Below we will see that these domain walls within domain walls indeed carry the $SU(2)$ quantum numbers despite of the fact that the global symmetry of the model is $SO(3)$. 

If the second scenario is realized, the $\phi$-excitations eventually condense on the wall (this cannot happen in the semiclassical limit), and restore the $\CC$-symmetry. However as we shall see the domain wall does not reduce to a trivial theory, as this would be inconsistent with the anomaly, but instead carries massless $SU(2)$ excitations. This can almost be seen from the semi-classical picture, as the excitations which condense on the wall to restore the $\CC$-symmetry  carry $SU(2)$ quantum numbers. 

\subsection{The lattice description}

\begin{figure}[tbp] 
   \centering
   \includegraphics[width=.5\textwidth]{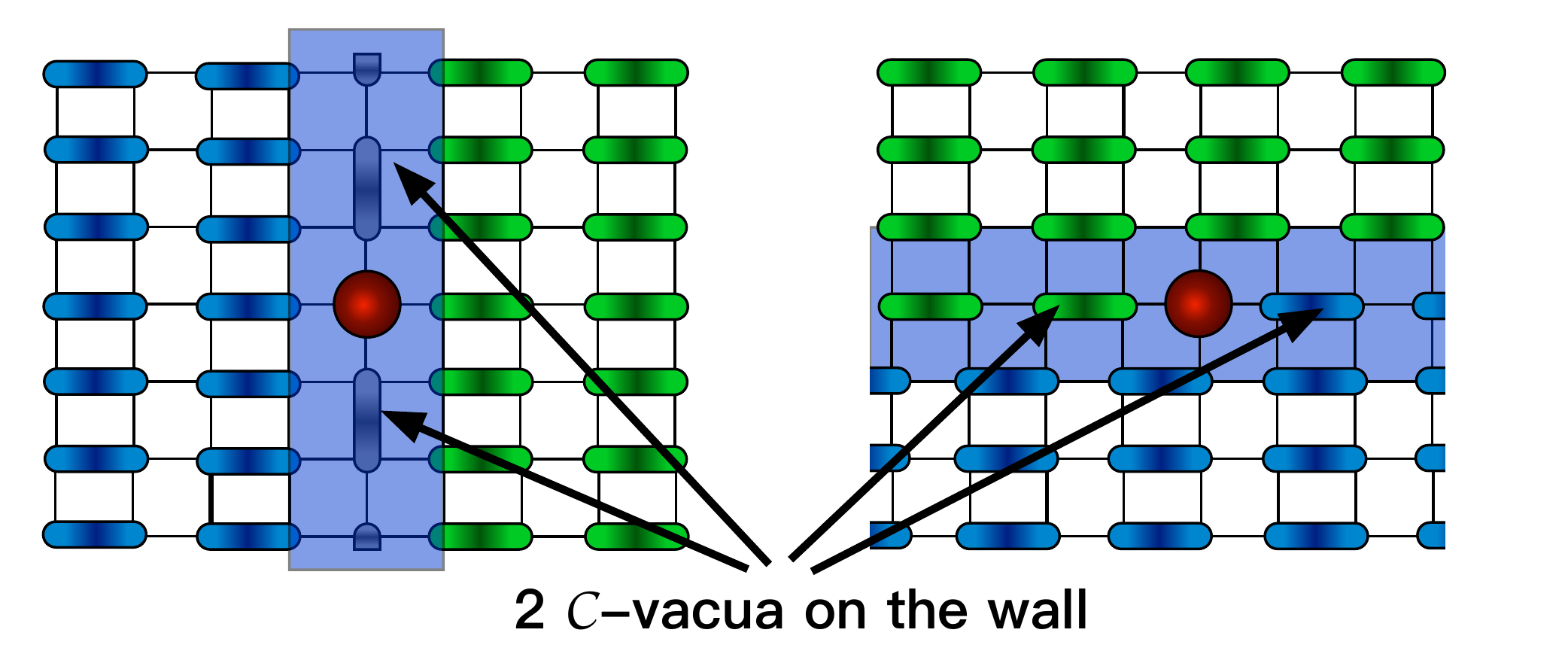} 
   \caption{A lattice depiction of $\Z_2$ VBS domain walls which break the $\CC$-symmetry.}
   \label{fig:DW_latt}
\end{figure}

Another way to get to the same conclusions is to think of the lattice description of the $\Z_2$ VBS ground state. In fact there are two realizations of the setup we discuss in this paper.  One concerns spin-$1/2$ anti-ferromagnet on the rectangular lattice, where the simplest picture of the VBS is in terms of spin singlets forming on nearest-neighbor links, and arranging themselves into horizontal formation. The domain walls are shown in Fig. \ref{fig:DW_latt}, and they support deconfined spinons, as discussed in \cite{Sulejmanpasic:2016uwq}.\footnote{We would also like to thank C. Wang for discussions of this interpretation.} Little thought reveals that the domain walls break up/down translational symmetry modulo $2$, which is to be identified with $\CC$-symmetry, and the spinon (depicted in red) acts as a domain wall inside the domain wall interpolating between the two $\CC$-vacua. Alternatively the spinons can condense on the wall, breaking the singlet dimers and restoring the $\CC$-symmetry. In this case, as we shall see, the theory must be gapless and consistent with the $SU(2)_1$ WZW theory.

\begin{figure}[tbp] 
   \centering
   \includegraphics[width=.5\textwidth]{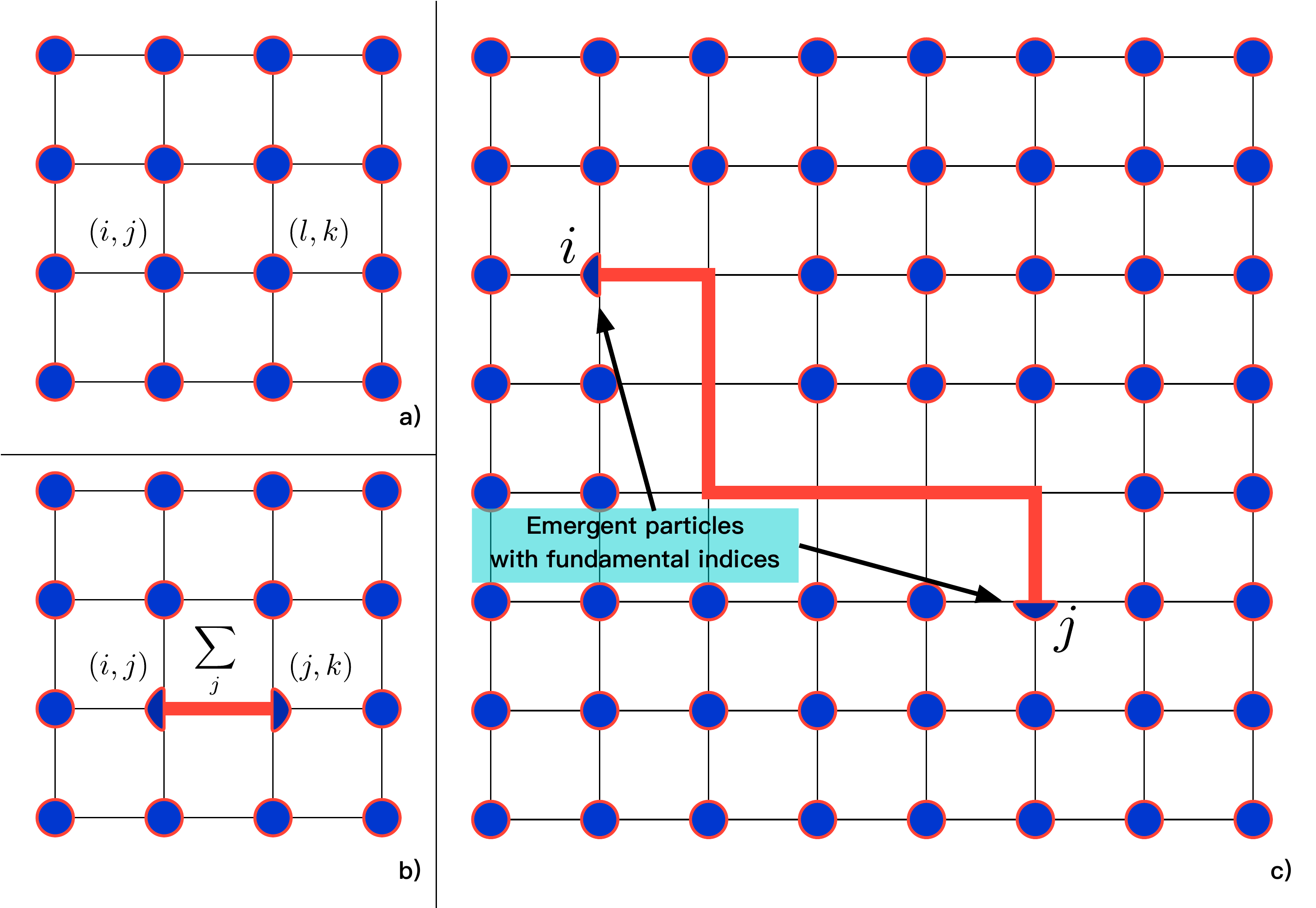} 
   \caption{A lattice depiction of spin-$1$ system. a) The product state: circles label spin-$1$ states on sites, labeled by two fundamental indices. b) The formation of the valence bond between two sites is a contraction of one of the fundamental indices on each site. c) A formation of a long valence bond with fundamental indices at the end.}
   \label{fig:S1_higgs}
\end{figure}

Another scenario concerns a spin-$1$ anti-ferromagnet on a square lattice. Remarkably the effective low energy theory still allows for emergent spinons carrying spin-$1/2$. To see this consider a product state of spin-$1$ triplet, as in Fig.~\ref{fig:S1_higgs}a. Two sites were labeled with two symmetrized fundamental indices $(i,j)$ and $(l,k)$ respectively. Now if we set $l=j$ and sum over it, we end up with two indices $i$ and $k$ living on two separate lattice sites.  The contraction of indices between the sites we labeled with a red line in Fig.~\ref{fig:S1_higgs}b. If we repeat the procedure, we can obtain a state with spatially well separated indices $(i,j)$ connected by a red line -- often called an AKLT chain \cite{affleck1987,affleck1988} -- as in Fig.~\ref{fig:S1_higgs}c as in Fig.~\ref{fig:S1_higgs}c. Note that while such a state is still in the SO(3) representation, in agreement with the global symmetries of the model, its group theory indices are spatially separated, giving rise to two emergent spin-$1/2$ particles at each end.

\begin{figure}[tbp] 
   \centering
   \includegraphics[width=.5\textwidth]{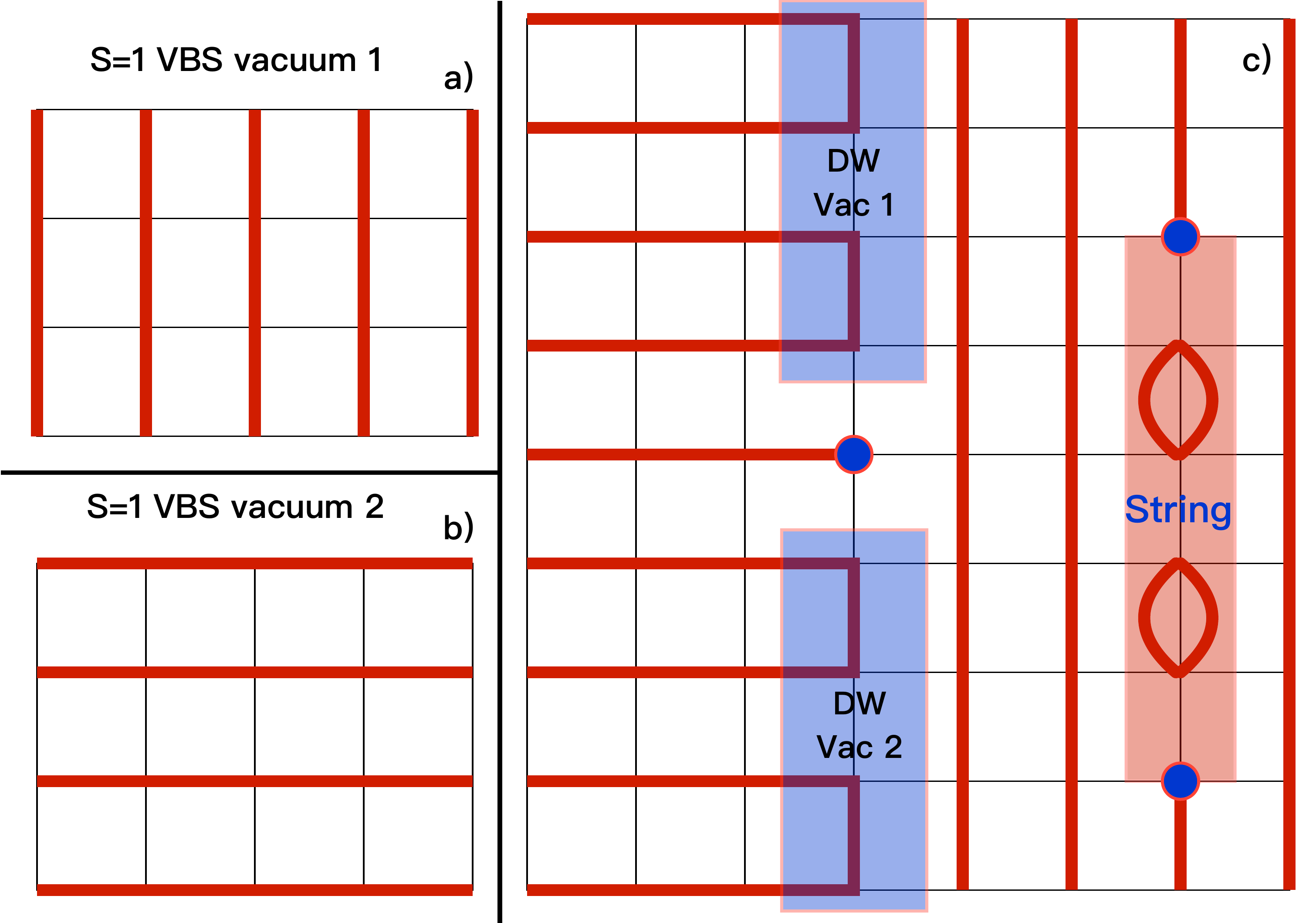} 
   \caption{a,b) Two vacua of the spin-$1$ VBS phase. c) A domain wall hosts free spin-$1/2$ particles (labeled by a blue circle), which are confined in the bulk.}
   \label{fig:S1_VBS}
\end{figure}

Nevertheless such a state is typically unfavored, both in the N\'eel phase, which is depicted in Fig.~\ref{fig:S1_higgs}c), where it would disintegrate easily into the vacuum by disintegrating the red bonds between the sites $i$ and $j$, reducing to the product state of spin-$1$ particles, while in the VBS phase\footnote{We assume here that there is a Hamiltonian which realizes such states as ground-states. A model which realizes a ground-state with these properties should be possible to construct by adding "Q terms" like in the famous J-Q model used in many numerical simulations (see \cite{Kaul:2012zv} for a review). }, depicted in Fig.~\ref{fig:S1_VBS}a,b such objects are typically confined, and do not exist as a part of the spectrum. Remarkably however, the domain wall between the two vacua of the VBS phase (depicted in Fig.~\ref{fig:S1_VBS}a,b) hosts deconfined spin-$1/2$ excitations. This is depicted in Fig.~\ref{fig:S1_VBS}c.

We will now see how these scenarios are consistent with and further constrained by the 't Hooft anomalies in the problem. 

\subsection{The 't Hooft Anomalies}

\begin{figure}[htbp] 
   \centering
   \includegraphics[width=.5\textwidth]{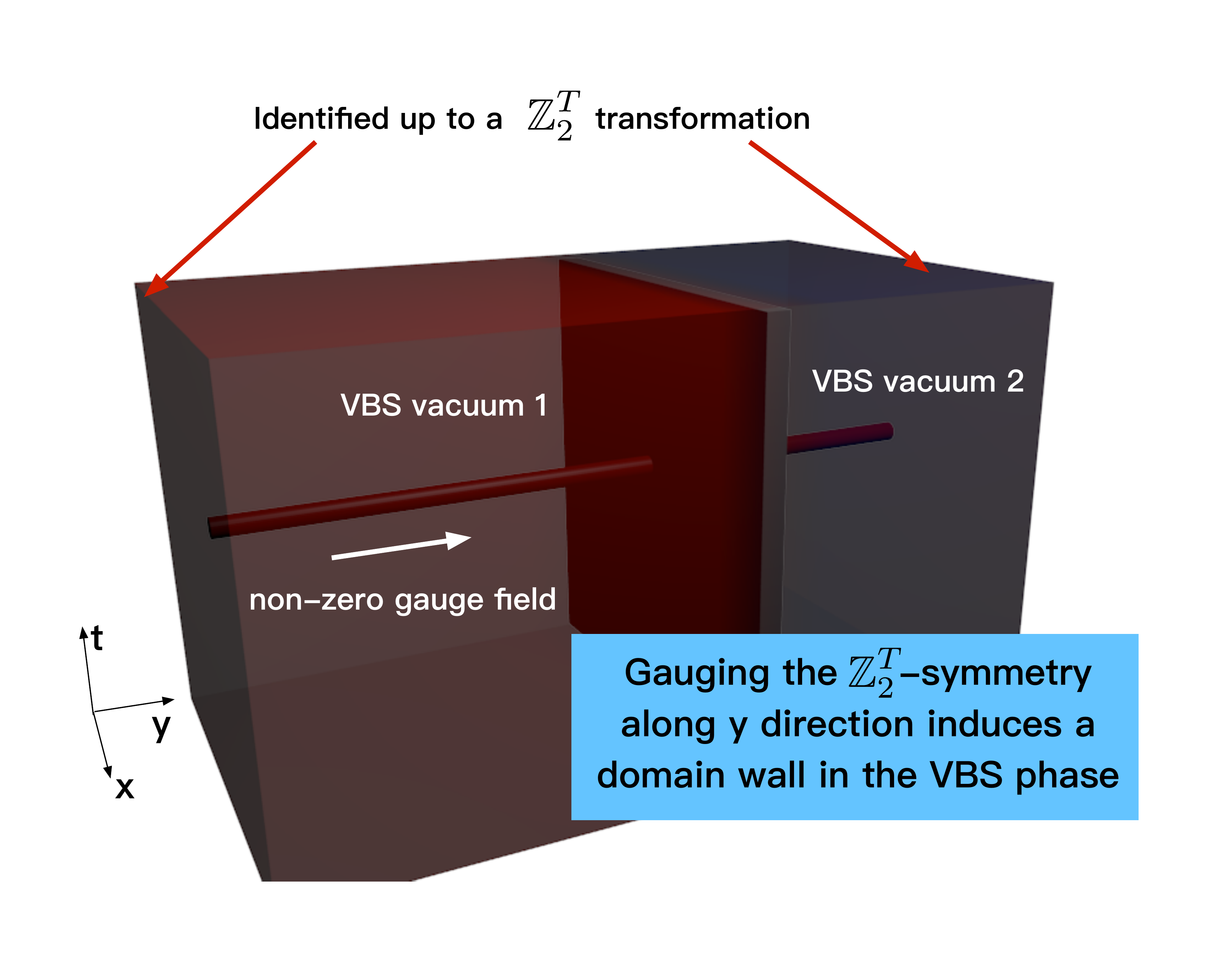} 
   \caption{A depiction of the space-time of the quantum magnet. By setting the $\Z_2$-topological gauge field $A$ to a constant along the $x$ direction, the VBS domain wall is induced. }
   \label{fig:Z2_gauging}
\end{figure}

We will now discuss the 't Hooft anomalies in this system and how they affect the domain wall physics (for a general discussion and derivation of the relevant 't Hooft anomalies see~\cite{Komargodski:2017dmc}). We will consider the  theory on an Euclidean $3$-torus with radii much larger than the inverse mass-gap of the theory, and label the directions as $x,y$ and $t$. Consider coupling the topological current $j=\star da$ to the 1-form gauge field we call $A$. Further since we assume only $\Z_2\subset U(1)_T$ is intact, this gauge field is a $\Z_2$ gauge field, so that $\int_\gamma A=0,\pi\mod 2\pi$, where $\gamma$ is a nontrivial cycle. The coupling of $A$ to the current $\star da$ is given by a term
\be\label{eq:min_coupling}
\frac{i}{2\pi}\int A\wedge F_{a}
\ee
in the action, where $F_a=da$ is the curvature 2-form. Since we define the VBS as the phase where $\Z_2$-topological is spontaneously broken, setting $A$ to a constant along, say the compact $y$-direction, so that $\oint_{\gamma_y}A=\pi$ induces a domain-wall interpolating between the two $\Z_2$ vacua (see Fig. \ref{fig:Z2_gauging}). But then the term above becomes
\be\label{eq:haldene_term}
\frac{i}{2}\int_{\Gamma} F_a
\ee
\begin{figure}[htbp] 
   \centering
   \includegraphics[width=.5\textwidth]{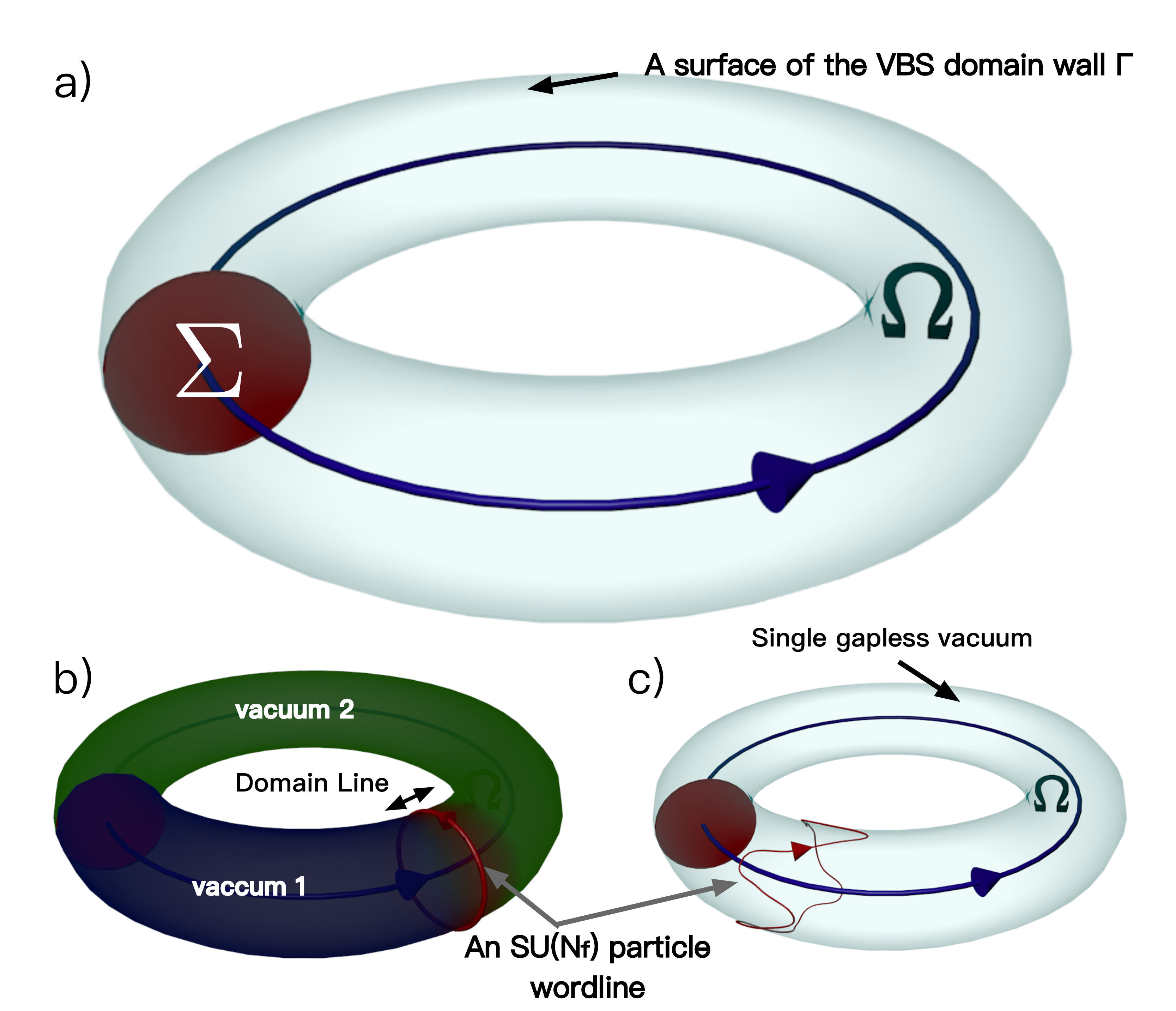} 
   \caption{A representation of the domain-wall surface on a torus a) and the two possible scenarios consistent with the 't Hooft anomaly inflow when the charge-conjugation gauge field is turned on along the cycle $\gamma_x$. This can be thought of as implementing a $C$ transformation along the disk $\Sigma$ on which an integral $\int_{\Sigma} B$ is not gauge invariant and requires a particle world-line in the fundamental $SU(N_f)$ representation wrapping along the minor cycle of the torus to cancel the gauge variation. }
   \label{fig:torus}
\end{figure}
over the 2-cycle $\Gamma$ complementary to $\gamma_y$. $\Gamma$ can be visualized as the surface of a torus, like in Fig. \ref{fig:torus}a.  Since the flux of $F_a$ along any cycle can only be an integer multiple of $2\pi$, the charge conjugation symmetry remains a symmetry. However if we promote the $U(1)$ gauge connection to an $U(N_f)$ gauge connection $\tilde a$, we must promote of $F_{a}\rightarrow \frac{1}{N_f}\Tr F_{\tilde a}$, and its integral is no longer quantized in multiples  of $2\pi$, but is now quantized in units of $\frac{2\pi}{N_f}$, which destroys the charge conjugation symmetry if $N_f>1$\footnote{If $N_f=1$ there is no anomaly. However there should still be an Ising transition between the C-broken phase and trivially gapped phase on the domain wall, discussed in \cite{Komargodski:2017dmc} (see also the discussion at the end of this section).}. In fact we can see that as we perform the $\CC$-transformation the relevant term shifts as (modulo an integer times $2\pi$)
\be
\frac{i}{2}\int_{\Gamma}\frac{1}{N_f}\tr{F_{\tilde a}}\rightarrow \frac{i}{2}\int_{\Gamma}\frac{1}{N_f}\tr F_{\tilde a}+i\int_{\Gamma} B
\ee
where $B$ is a 2-form $\Z_{N_f}$ gauge field so that its integral is given by\footnote{The gauge fields of $U(N_f)$ should be separated into the dynamical part $U(1)$ and the non-dynamical part $PSU(N_f)$, which can be thought of as $SU(N_f)$ with its center symmetry $\Z_{N_f}$ gauged away by a two-form gauge field $B$.} $\int B=\frac{2\pi k}{N_f}\mod 2\pi, k\in\mathbb Z$. For $N_f=2$ we can identify $B=\pi w_2(SO(3))$, where $w_2$ is the second Stiefel-Whitney class of the $SO(3)$ bundle. Note that the integral of the $B$-field is only well-defined up to an overall $2\pi$ integer multiple. 

To restore charge conjugation symmetry we may try to add a local term in the the action $p\int_{\gamma} B$, where $p$ is an integer.  The transformation will then be (see~\cite{Komargodski:2017dmc} for more details)
\begin{multline}
\frac{i}{2}\int_{\Gamma}\frac{1}{N_f}\tr F_{\tilde a}+ip\int_{\Gamma}B\rightarrow \\\rightarrow \frac{i}{2}\int_{\Gamma}\frac{1}{N_f}\tr F_{\tilde a}+i(1-p)\int_{\Gamma} B
\end{multline}
If $1-p=p\mod N_f$ then the theory has recovered its $\CC$-invariance. However this cannot be achieved with integer $p$ for even $N_f$ (for odd $N_f$ a more lengthy argument is needed, with the same conclusion. See \cite{Komargodski:2017dmc}.). The arguments above show that the domain wall theory has a mixed charge-conjugation $PSU(N_f)$ anomaly\footnote{The theory also has a $\Z_2\times PSU(N_f)$ mixed anomaly, indicating that the ground state must necessarily be non-trivial. One can see this from noticing that promoting $F_{a}\rightarrow \tr F_{\tilde a}/N_f$ in \eqref{eq:min_coupling} destroys the gauge invariance of $\Z_2$ gauge field $A$.}. For the case of $N_f=2$ one can describe this anomaly by \cite{Komargodski:2017dmc}
\be \label{wthree}\pi i \int_3 w_3(O(3))~,\ee 
i.e. the third Stiefel-Whitney class of $O(3)=\CC\times SO(3)$, integrated over some auxiliary three-dimensional space that ends on our domain wall  (see Fig.~\ref{fig:torus}). In other words the domain wall has an anomaly between $SO(3)$ symmetry and charge conjugation, and as a consequence cannot be trivial. One can think about the term~\eqref{wthree} as an integral of the product between the second Stiefel-Whitney class $w_2(SO(3))$ and the charge conjugation gauge field. 

To see this, consider introducing the bulk term\footnote{Note that while $dB=0$ on the surface of the domain wall, this is not necessarily the case in the auxiliary dimension. This is akin to considering a flat, but nontrivial, $U(1)$ connection on a circle extended to a disk which indicates the presence of a magnetic field somewhere in the bulk.}
\be
\frac{i}{2}\int_{\Omega} dB\;.
\ee
such that the boundary of $\Omega$ is $\Gamma$ --- the worldsheet of the domain wall as in Fig.~\ref{fig:torus}.
Let us now gauge the $\CC$-symmetry by introducing a 1-form $\Z_2$ gauge field $C$ with $e^{i\oint C}=\pm 1$. When $e^{i\oint_{\gamma}C }=-1$ all the fields transforming under $\CC$ must experience a jump as $\gamma$ is traversed. If we now pick a cycle traversing the $x$-direction $\gamma_x$ and select $C$ such that $e^{i\oint_{\gamma_x}C }=-1$, the above term becomes
\be
i\int_{\Sigma}B
\ee
where $\Sigma$ is a surface in the bulk of $\Omega$ which ends on the domain-wall $\Gamma$. 

However, as we commented early on, this integral is not well defined unless there is an object on the boundary which cancels the gauge variation of $B\rightarrow B+d\lambda$. A natural object which accomplishes this is an $SU(N_f)$ fundamental Wilson loop. Hence we must have a fundamental Wilson loop -- a particle world-line carrying $SU(N_f)$ quantum numbers -- winding somewhere on the surface $\Gamma$ (see Fig. \ref{fig:torus}b,c), which we can think of as the surface of the domain-wall.

These facts are rather manifest in the semi-classical description, as we explained above. However we see here that the anomaly is a powerful tool that allows us to make some statements also away from the semi-classical limit.

\begin{itemize}[leftmargin=10pt]
\item If the charge-conjugation is spontaneously broken on the domain wall $\Gamma$, the constant $C$-field along $\gamma_x$ will induce a domain-wall (inside the original domain-wall) somewhere on $\Gamma$ (see Fig.~\ref{fig:torus}b). The anomaly then tells us that this domain-wall must support a  particle world-line in the $SU(N_f)$ fundamental representation. So a massive particle excitation carrying quantum numbers of the fundamental $SU(N_f)$ representation inside the domain wall exists, and they are identified as domain-walls of the $\CC$-broken domain wall theory. This is true whether or not we are in the semi-classical limit.
\item If charge conjugation on the domain wall \emph{is not} spontaneously broken (for example, it could be restored away from the semi-classical limit), the $C$-field will not change the vacuum as $\gamma_x$ is traversed. However the anomaly matching \emph{insists} that there is a particle wrapping around $\gamma$ in order to compensate the gauge variation of $B$. Hence we must be in a phase where there are massless $SU(N_f)$ excitations. 
\end{itemize}

Another possibility which can in principle be realized is that the domain wall theory restores all the symmetries but has a nontrivial TQFT, e.g. see the analysis along the lines of~\cite{Seiberg:2016rsg}. We will assume that this does not happen.

In all of the cases above, however, the anomaly guarantees that on the domain wall excitations exist which carry fundamental $SU(N_f)$ quantum numbers, which do not exist in the bulk (as they are confined and the local operators sit in $PSU(N_f)$ representations). Hence we conclude that the domain-wall theory is always deconfining. 

When $\CC$-symmetry is restored, the domain wall can become gapless. A natural candidate for the theory on the domain-wall in this case is the $SU(2)_1$ Wess-Zumino-Witten model in 1+1D.  This particular conformal field theory matches the required mixed anomaly because it is also the universality class of the $O(3)$ model at $\theta=\pi$, which has the same anomaly~\eqref{wthree}. This nicely fits with the numerical measurements in the J-Q model \cite{Sulejmanpasic:2016uwq}. and the fact that in the lattice construction of the $\Z_2$ VBS phase the domain wall closely resembles the spin-$1/2$ chain which was also suggested in \cite{Sulejmanpasic:2016uwq}.

We note that it is also possible that both $\CC$ and $\Z_2$ are spontaneously broken. This can be achieved, for instance, if we add to~\eqref{eq:sigma_lag} a term like $\cos(4\varphi)$ with a sufficiently large coefficient. In that case there are four vacua and two distinct, trivial domain walls. This is also consistent with the anomaly inflow arguments.

We briefly comment on the case of $\Z_4$ VBS, where 4-monopole events are allowed. The symmetry is spontaneously broken in the VBS phase, and, hence, has stable domain walls. Semi-classically the theory is described by a scalar field and potential of the form $-\cos(4\varphi)$. The domain walls which interpolate between $\varphi:2\pi k/4\rightarrow 2\pi (k+1)/4$ do not host deconfined spinons. There is no anomaly inflow on these walls and the domain wall is expected to be trivial.

On the other hand, the intersection of four domain walls that are related cyclically by the $\Z_4$ topological symmetry is nontrivial. The intersection hosts a spinon~\cite{levin2004deconfined}. Interestingly this is guaranteed by anomaly inflow. To see this, let us impose a vortex in the $\Z_4$ gauge field $A$ for $U(1)_T$, so that $dA=2\pi\delta(\mathcal C)$, where $\delta(\mathcal C)$ is a 2D $\delta$-function along the world-line of the vortex. By gauging the $PSU(N)$ symmetry we have that $\frac{i}{2\pi}\int_{\mathcal M} A\wedge \tr{F}/N_f$ is not gauge invariant under the gauge transformations of $A$ and needs to be supplemented by a bulk term\footnote{Note that this indicates a mixed anomaly between the topological symmetry and $SO(3)$, and no $C$ symmetry appears. The $C$ symmetry is not necessary for the present discussion.} $\frac{i}{2\pi}\int_{\Theta} A\wedge dB$, where $\partial \Theta=\mathcal M$. Partially integrating this term we get $\frac{i}{2\pi}\int_{\Theta}dAB=i\int_{\mathcal C} B$, hence the vortex carries a projective $PSU(N_f)$ representation, which is really just the fundamental representation of $SU(N_f)$. For $N_f=2$ this is just a qubit.

Finally, the entire discussion is easily adapted to $4D$ where the relevant topological symmetry is a $1$-form symmetry. We consider a $U(1)$ gauge theory with $N_f$ electrically charged particles and we assume that there are only even magnetic monopoles. The system then has a magnetic $\Z_2$ symmetry, under which the fundamental 't Hooft line carries charge 1, and a $PSU(N_f)$ symmetry acting on the charge particles, exactly like before. 

Let us assume that our even charge monopole particles condense. The $1$-form $\Z_2$ topological symmetry is spontaneously broken, allowing for stable Abrikosov-Nielsen-Olsen {\it electric} strings, carrying half the flux of the fundamental electric charge. The low energy effective theory is a $\Z_2$ gauge theory, which has essentially a line and a surface which can link each other. The surface is the string with half electric charge and the line is the 't Hooft line with magnetic charge 1.
The strings are stable because the dynamical charges are unit charges, and half-electric fluxes cannot end on them. 
Let us couple the $\Z_2$-topological 1-form symmetry to a background 2-form gauge field $A$. The coupling looks the same as~\eqref{eq:min_coupling} only it is now a four-dimensional integral. Setting $\oint_{\gamma}A=\pi\mod 2\pi$, where $\gamma$ is a 2-cycle now, induces an ANO vortex along the complementary 2-cycle $\Gamma$, reducing to \eqref{eq:haldene_term}.

We again distinguish two cases: $N_f=1$ and $N_f>1$. The $N_f=1$ model on the vortex can be therefore viewed as 1+1 dimensional QED at $\theta=\pi$ with dynamical unit charges. This model has no anomaly. It was analyzed in detail in~\cite{Komargodski:2017dmc} (and see references therein). In the semi-classical limit where the electric charges are heavy we have two ground states and charge conjugation (alternatively, time reversal) is spontaneously broken. We may ponder what happens as the electric charges are decreased in mass. The pure 1+1 dimensional model is dual to the Ising model and so the model becomes critical and then switches to a trivial ground state (for which there is no obstruction since there is no anomaly) in which the electric charges are condensed. It would be interesting to see if the domain wall in this theory undergoes similar dynamics, i.e. develops an Ising fixed point before the bulk becomes gapless. 

On the other hand if $N_f>1$, the vortex has an anomaly between the $\CC$ and $PSU(N_f)$, so it can either break $\CC$-symmetry or be gapless, just like before. The anomaly again insists that the wall supports deconfined $SU(N_f)$ excitations, by the same arguments as before.

\section{Easy-Plane/Axis $\C\P^{1}$ Model}\label{sec:easy-plane}

The easy-plane and easy-axis models are spin models where the $O(3)$ spin-symmetry is reduced to the $\Z_2\times O(2)$ subgroup and they play an important role in the physics of materials. The regime of easy-plane and easy-axis is differentiated by whether the N\'eel-vector prefers the plane on the equator or the poles of the Bloch sphere. We will often refer to both of them as ``easy-plane'' for simplicity. 

These models are of great interest in quantum magnets. They played an important role in establishing the Haldane conjecture \cite{Haldane:1982rj,Haldane:1983ru}, but are also more realistic as exact $O(3)$ spin-symmetry is unlikely to occur in realistic systems. Therefore they are interesting to study (see e.g. \cite{Motrunich2017,Wang:2017txt} for recent references).

We describe here an 't Hooft anomaly in the 2+1 dimensional Easy-Plane $\C\P^{1}$ model and we outline its consequences for the bulk phases of the model as well as the domain wall phases\foot{We are grateful to O. Motrunich who prompted us to consider this problem.}. First, let us define the model in terms of the continuum 2+1 dimensional theory 
\begin{multline}\label{LagEPCP}
{\cal L}=-{\frac{1}{4e^2} }|da|^2+\sum_i |D_a\phi_i|^2\\+m_1^2|\phi_1|^2+m_2^2|\phi_2|^2+\lambda (|\phi_1|^2-|\phi_2|^2)^2~.
\end{multline}
Both $\phi_{1,2}$ have charge $+1$ under the gauge field $a$. Being for a moment careless about various discrete factors, the internal symmetries of the model are $U(1)\times U(1)_T$, where $U(1)$ acts as $\phi_1\to e^{i\alpha}\phi_1$, $\phi_2\to e^{-i\alpha}\phi_2$ and $U(1)_T$ is the standard topological symmetry generated by the charge ${1\over 2\pi}\int_2 da$.

\begin{figure}[htbp] 
   \centering
   \includegraphics[width=.5\textwidth]{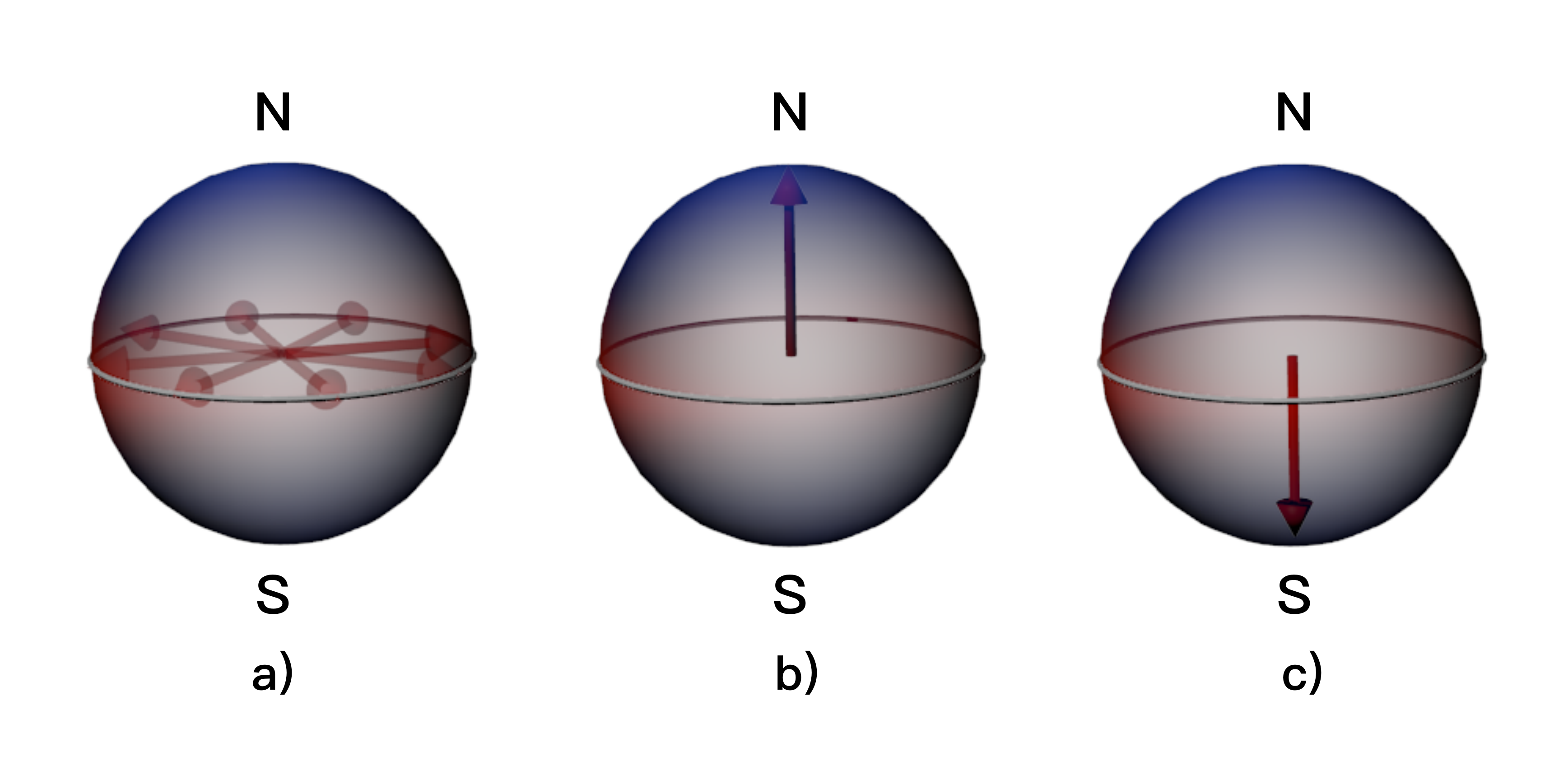} 
   \caption{a) The preferred orientation of the $O(3)$ order parameter $\vec n=\sum_{i,j=1,2}\phi^*_i (\vec\sigma)_{ij} \phi_j$ (red arrow), when $\lambda$ in \eqref{LagEPCP} is positive. b) and c) The preferred orientation of $\vec n$ when $\lambda$ is negative. The $\Z_2$ exchange symmetry corresponds to the exchange of north and south poles.}
   \label{fig:easy-axis}
\end{figure}

 First, let us do some semi-classical analysis in order to get some intuition for the possible phases of the model. Let us first consider a phase where $\lambda>0$, so that the vacuum prefers that $|\phi_1|=|\phi_2|$. The model therefore prefers the equator of the $O(3)$ order parameter (see Fig. \ref{fig:easy-axis}a). If both $m_1^2$,$m_2^2$ are large and positive, the dynamics is as before. The model breaks spontaneously the discrete subgroup of $U(1)_T$ that is preserved by the monopole operators and $U(1)$ is unbroken. If both are large and negative, then there is again a massless Nambu-Goldstone boson corresponding to a broken $U(1)$ but now $U(1)_T$ is preserved. If however one of the masses squared is large and negative and the other is large and positive then the model is in a disordered phase, where the vacuum is gapped and trivial. (In some limit, where the positive mass squared is much larger in absolute value than the negative mass squared, we can think about this trivial vacuum as being the trivial vacuum of the $O(2)$ Wilson-Fisher model by particle-vortex duality~\cite{Peskin:1977kp,Dasgupta:1981zz}). This analysis makes it evident that a disordered phase exists and hence the model cannot have a 't Hooft anomaly. 
But this analysis also suggests that if we include one more discrete symmetry that pins $m_1^2=m_2^2=m^2$, then a 't Hooft anomaly may exist.  Such a symmetry is given by the $\Z_2$ exchange symmetry $\phi_1\leftrightarrow \phi_2$.

Furthermore, if  $m^2$ is large and positive, and if the even monopoles are added, the system  is in the $2$-state VBS phase which supports domain walls with a semi-classical description just as in the limit where there is the full $O(3)$ symmetry we discussed before. 

\begin{figure}[tbp] 
   \centering
   \includegraphics[width=.5\textwidth]{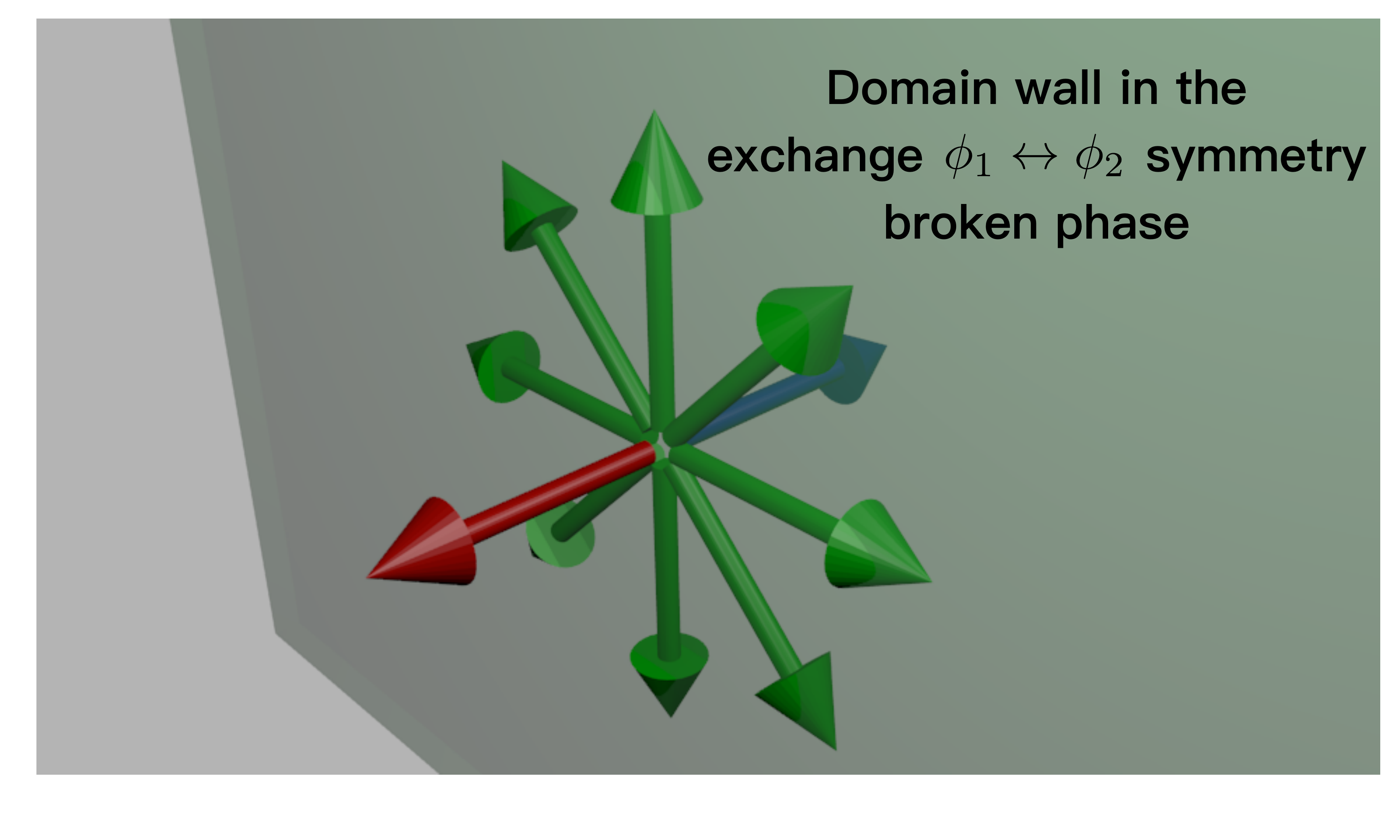} 
   \caption{A representation of the domain wall of the $\Z_2$-exchange symmetry broken phase. On the one side of the domain wall we must have the $O(3)$ order parameter pointing  north (the red arrow) and on the other south (the blue arrow). The monopole--anti-monopole is linearly confined in the bulk, but it is only logarithmically confined on the domain wall  as a monopole becomes a vortex on the domain wall (pictured). The system can therefore undergo the BKT transition on the domain wall by percolating vortex-monopoles. If only even monopoles are allowed in the bulk, correspondingly even vortices (winding modes) will be allowed on the domain wall. }
   \label{fig:vortex_monopole}
\end{figure}

Alternatively we can have a phase where $\lambda$ is negative and sufficiently large. The vacuum spontaneously breaks the $\Z_2$ exchange symmetry $\phi_1\leftrightarrow \phi_2$. The domain wall in this case is an $O(2)$ sigma model (i.e. the $c=1$ model at large radius)  in 1+1 dimensions, with vortices (winding modes added to the Lagrangian). That this is true can be seen by solving for the domain wall, and integrating over the direction perpendicular to it. Note that the vortices on such a domain wall are trapped monopoles (i.e. hedgehogs in the $O(3)$ order parameter) in the full theory (see Fig. \ref{fig:vortex_monopole}). This is a curious case because the domain wall is exactly massless even when the bulk is gapped and the physics is semi-classical.  

It is possible that as the model is deformed towards the transition to the $\Z_2$ VBS by dialing $\lambda$, (double charged) monopole-vortices condense on the wall before the transition in the bulk occurs. Furthermore this possibly occurs even when there is no topological symmetry (i.e. when unit-charge monopoles are allowed), and no anomaly. In other words, as $\lambda$ is increased,  the domain wall may become trivial before the vacuum undergoes a transition to being trivially gapped (and the domain wall disappears).

We will see that non-trivial domain walls are required by anomaly inflow coming from the bulk 't Hooft anomalies involving the topological symmetry, $\Z_2$ exchange symmetry, the $U(1)$ flavor symmetry and the $\CC$ charge conjugation symmetry.

To demonstrate this, let us first couple the system to a background gauge field $A$ for the $U(1)_T$ symmetry. This is done via the standard coupling 
\begin{equation}\label{Aa}{1\over 2\pi}\int A\wedge da~.\end{equation}
This coupling is perfectly gauge invariant under $A$ and $a$ gauge transformations.
Now let us note that we were somewhat imprecise about the action of the $U(1)$ symmetry, $\phi_1\to e^{i\alpha}\phi_1$, $\phi_2\to e^{-i\alpha}\phi_2$. At $\alpha=\pi$ this coincides with a gauge transformation. In other words, the scalars $\phi_{1,2}$ transform under the total gauge symmetry $\frac{U(1)_a\times U(1)}{ \Z_2}$. Therefore, in the presence of background gauge field $B$ for the $U(1)$ symmetry, the fluxes of $da$ can be half-integer because the holonomies can be unwound through the gauge field $B$. These fluxes are correlated, so 
$$\int_2 da= \int_2 dB \mod 2\pi~.$$
Due to these possible half-integer fluxes, the coupling~\eqref{Aa} may now break the gauge invariance under $A\rightarrow A+d\lambda$. We can fix it with a properly quantized Chern-Simons counter-term (Chern-Simons counter-terms must always be properly quantized~\cite{Closset:2012vg,Closset:2012vp}) for the background gauge fields by adding another term to~\eqref{Aa}: 
\begin{equation}\label{AaB}{1\over 2\pi}\int A\wedge da+{1\over 2\pi}\int A\wedge dB~.\end{equation}
Since the half-integer part is correlated, and since now $A$ couples to $da+dB$, the action is again perfectly gauge invariant under $A$ and $B$ gauge transformations.

So far what we have seen is that the partition function, as a functional of $A,B$ gauge fields, can be rendered perfectly gauge invariant by appropriate choices of the counter-terms. Therefore there is no 't Hooft anomaly that involves just the symmetries $U(1)\times U(1)_T$. This is of course nicely consistent with the semi-classical analysis, which exhibited a disordered phase in parts of the parameter space. 

Now let us imagine that we require the $\Z_2$ symmetry $\phi_1\leftrightarrow \phi_2$. This in particular requires that $m_1^2=m_2^2$ in the Lagrangian~\eqref{LagEPCP}. This $\Z_2$ commutes with $U(1)_T$ but it does not commute with $U(1)$; it is now extended to $O(2)$. Therefore, under the action of the $\Z_2$ symmetry $A\to A$ and $B\to -B$. Also the dynamical gauge field $a$ is invariant under this symmetry. 
The counter-term that we have added in~\eqref{AaB}, ${1\over 2\pi}\int A\wedge dB$, breaks this $\Z_2$ symmetry explicitly (see also \cite{Wang:2017txt}). The statement therefore  is that we cannot preserve both $U(1)_T$ gauge invariance in the presence of nontrivial $B$ and preserve the $\Z_2$ symmetry. This is therefore an anomaly that involves all these three symmetries. As usual in such situations, we could have added a four-dimensional bulk counter-term that preserves all the symmetries, including the $\Z_2$ symmetry. Such a term would be $\frac{i}{2\pi}\int_4 dA\wedge dB$.   

If we introduce a $\Z_2$ valued gauge field $D$ that couples to our $\phi_1\leftrightarrow \phi_2$ symmetry, we could write the anomaly polynomial in terms of an auxiliary 3+1 dimensional bulk  as
\begin{equation}\label{easyanomaly}{1\over 2\pi}\int_4 D \wedge A \wedge dB~. \end{equation}
This writing is however schematic; it is meant to convey the symmetries that participate in the anomaly. It is only schematic because $B$ is not invariant under the $\Z_2$ that couples to the gauge field $D$. 
In addition, so far we have ignored charge conjugation symmetry. (It will be  important below.)

As usual, the existence of the anomaly~\eqref{easyanomaly} immediately implies that the model cannot be disordered (even far from the semi-classical regime) by perturbations that preserve all the symmetries.
As in our analysis of the N\'eel-VBS transition, let us now imagine that we add monopole operators. We may assume first that they are divisible by $4$, so we have a $\Z_4\subset U(1)_T$ unbroken symmetry. The domain walls in the confining phase are trivial but their clockwise intersection is nontrivial. We can think about it as turning a unit flux for $dA$. One finds that the core supports the anomaly polynomial (schematically) $\int_2 D\wedge B$, in other words, the core supports a projective $O(2)$ representation. Therefore, the core is doubly degenerate (i.e. there is a qubit at the core). This can be of course associated with the the particles $\phi_1,\phi_2$ sitting at the core. The Hilbert space is in a representation of $Pin(2)$ and the projective Hilbert space would transform under $O(2)$, as necessary. In fact the degree of freedom at the core is dual to a free complex fermion~\cite{Gomis}.  

Let us now assume that we add even monopoles to the Lagrangian. The $U(1)_T$ symmetry is now broken to $\Z_2$. The domain wall in the confining phase is now nontrivial. In order to understand its anomaly polynomial we have to reinstate the charge conjugation symmetry. If we denote the associated gauge field by $C$ there is now a new term in the anomaly polynomial, given by $$\int A \wedge C \wedge w_2(O(2))~,$$ where the $O(2)$ consists of the $SO(2)$ gauge field $B$ and the $\Z_2$ gauge field $D$.\footnote{One can view this as a reduction of the anomaly $\int_4 A\wedge w_3(O(3))$ of~\cite{Komargodski:2017dmc} (which is valid when the topological symmetry is broken to $\Z_2$) in the following way:  First, ignoring charge conjugation, the anomaly reduces reduce the anomaly to $\int_4 dA\wedge w_2(SO(3))$. Next, we observe that if we only preserved $SO(2)\subset SO(3)$ then the anomaly would disappear completely. This is the statement we derived above, that there is no anomaly in the $U(1)\times U(1)_T$ model. However, imagine that we preserve $O(2)\subset SO(3)$. The additional generator is our $\phi_1\leftrightarrow\phi_2$ symmetry. Then, the anomaly does not disappear and we remain with $\int_4 dA\wedge B \wedge C$, which is exactly what we obtained in~\eqref{easyanomaly}.  In the presence of a nontrivial charge conjugation gauge field $C$, the anomaly $\int_4 A\wedge w_3(O(3))$ reduces to $\int_4 A\wedge C\wedge w_2(SO(3))$ which further reduces to $\int_4 A\wedge C\wedge w_2(O(2))$ upon breaking $SO(3)$ to $O(2)$.}  This is again written schematically because charge conjugation acts on $SO(2)$.

The domain wall therefore has $\Z_2\ltimes O(2)$ symmetry and 't Hooft anomaly, schematically written as 
\begin{equation}\label{anoDW}\int_3 C\wedge w_2(O(2))~.\end{equation}

Semi-classically the domain wall breaks the charge conjugation symmetry and the $O(2)$ symmetry is preserved. So we again have domain walls within domain walls, now carrying projective $O(2)$ representations. 

We can ask what happens when we depart from the semi-classical regime. The theory on the wall may again become critical while the bulk is still gapped. A natural 1+1 dimensional conformal field theory saturating the above anomaly is the compact $c=1$ boson. The symmetry $\Z_2\ltimes O(2)$ guarantees that we should not add to the Lagrangian momentum modes and also we should not add odd winding modes. Therefore the transition is of the Berezinskii-Kosterlitz-Thouless (BKT) type, where the semi-classical phase maps to the gapped phase with two degenerate ground states where the winding modes condense and the quantum phase could map to the unbroken phase with power-law correlation functions. However it is not exactly the BKT transition as there is $\Z_2$ symmetry breaking/restoration involved and both sides of the transition are ordered. 

A geometric way to understand this is to start from the $O(3)$ model at $\theta=\pi$ --- i.e. an effective model of a spin-1/2 chain which matches all the anomalies --- and imagine breaking the $O(3)$ symmetry  (e.g. by adding a potential) to $\Z_2\ltimes O(2)$. In the language of the Abelian Higgs model that corresponds to
\begin{multline}\label{LagEPCPDW}
{\cal L}=-{\frac{1}{4e^2} }|da|^2+\sum_i |D_a\phi_i|^2+\frac{i}{2} da\\+m^2|\phi_1|^2+m^2|\phi_2|^2+|\phi_1|^4+|\phi_2|^4~.
\end{multline}
This model has charge conjugation symmetry $a\to -a$, $\phi_i\to \phi_i^*$ when $\theta=0,\pi$, where we used the fact that $\int da\in 2\pi \Z$ and hence $\frac12 da$ is invariant when integrated over a closed two-dimensional manifold.
This model has the anomaly when $\theta=\pi$ \eqref{anoDW} which we can verify upon coupling the model to a background gauge field for the symmetries $\Z_2:\phi_1\leftrightarrow\phi_2$, $U(1): \phi_1\to e^{i\alpha}\phi_1$, $\phi_2\to e^{-i\alpha}\phi_2$. 

If $m^2$ is large and positive then we have two ground states related by charge conjugation symmetry. If $m^2$ is large and negative then we have a large circle of vacua with irrelevant even winding mode operators added to the Lagrangian. 
The model has vortices which classically have a $\Z_2$ degeneracy at the core, because either $\phi_1$ of $\phi_2$ can go to zero there (i.e. the $O(3)$ N\'eel vector vector pointing north or south at the center of the core. These vortices are sometime called \emph{merons}, and they carry a half-integer topological charge.) Quantum-mechanically, the degeneracy is lifted for generic $\theta\ne \pi$, as the two core configurations are summed over and there is only one vortex in the IR theory, so that the $\Z_2$ symmetry acts trivially on the vortex operators. The IR theory is therefore a critical 1+1D $O(2)$ model, or the $c=1$ compact boson, with momentum and vortex operators. This model has no anomaly as in the vortex percolating phase it has a trivially gapped state, and therefore cannot be the model we seek.  

At $\theta=\pi$, however, the single winding vortices cancel, as the sum over the two $\Z_2$ orientations inside the core interferes destructively. As a result, only even vortex modes are allowed and the model has a $\Z_2$ topological symmetry. The model therefore reduces to the $\Z_2\times O(2)$ 1+1D compact boson. Let us reduce to this regime by taking the extreme limit of $m^2\rightarrow -\infty$ in \eqref{LagEPCPDW}. The model essentially reduces to 
\be
\mathcal L_{eff}= |d\alpha_1+a|^2+|d\alpha_2+a|^2+\frac{i\theta}{2\pi}da\;.
\ee
Let us dualize the above Lagrangian, and obtain
\be
\mathcal L_{eff}^{dual}= |d\sigma_1|^2+|d\sigma_2|^2+i\frac{1}{2\pi}(\theta-\sigma_1-\sigma_2)da\;.
\ee
By integrating out the gauge field, we get that there is only one degree of freedom as $\sigma_1=-\sigma_2-\theta$. The operator $e^{in\sigma_1}$ is a $n$-vortex operator. Under the exchange symmetry it transforms as
\be
\Z_2: e^{in\sigma_1}\rightarrow e^{in\sigma_2}=e^{-in\sigma_1-in\theta}\;.
\ee
At $\theta=0,\pi$ we also have a charge-conjugation symmetry, which acts as
\be
\CC: e^{in\sigma_1}\rightarrow e^{-in\sigma_1}\;.
\ee
Consider the combination at $\theta=\pi$
\be
Z_{2}\CC: e^{in\sigma_1}\rightarrow (-1)^ne^{in\sigma_1},
\ee
which is the same as $\sigma_1\rightarrow \sigma_1+\pi$. Since this is an exact symmetry of the theory at $\theta=\pi$, the vortex potential must be of the form $\sum_{n}c_n \cos(2n\sigma_1)$.

Let us summarize: the limit $m^2$--- large and negative contains vortex operators which wind the global $U(1)\in SO(3)$ that acts on the two scalars oppositely. This vortex forces a half-instanton (i.e. meron) at its core, and hence couples as $\exp(\pm i\theta/2)$. The sign depends on the (classical) $Z_2$--exchange symmetry vacua at the core of the vortex, i.e. it depends on whether $\phi_1$ or $\phi_2$ vanished at the core. Quantum mechanically, the two vortex vacua are summed over, which causes single-vortices to interfere with each other at $\theta=\pi$. Hence only even dynamical vortices are allowed.

Note that in this model where only even winding modes are allowed in the action, the discrete anomaly which we derived starting from 2+1 dimensions becomes a discrete anomaly in 1+1 dimensions which can be viewed as a standard $\Z_2$ 't Hooft anomaly involving the axial and vector symmetries of the compact bosons.

\section{Conclusions and Summary}

In this work we have discussed domain walls in the VBS phase of quantum anti-ferromagnets in 2+1D. At low energies these systems are described by the Abelian-Higgs system with a scalar doublet and with monopole insertions. We were mostly concerned with the $\Z_2$-VBS phase, which is a gapped phase of percolating charge-2 monopoles with two nonequivalent vacuum states and a domain wall between them. The VBS vacuum further allows only spin-1 excitations in the bulk, while spin-1/2 excitations -- spinons -- are confined. 

We have shown that the domain wall carries a 't~Hooft anomaly between charge conjugation symmetry and $SO(3)$ global (spin) symmetry, and as a result either breaks the charge conjugation symmetry or is gapless. In both of these scenarios the domain wall theory supports deconfined spinons. The scenario in which the domain wall breaks charge conjugation symmetry is realized when the mass-squared of the Higgs doublet is large. We have argued that as the mass of the scalars is lowered the domain wall theory may undergoe a phase transition to the gapless WZW theory, while the bulk still remains gapped. This scenario is supported by the recent first-principle Monte Carlo simulations of the J-Q model \cite{Sulejmanpasic:2016uwq}. We also discussed the 3+1D Abelian-Higgs theory with charge-2 monopoles, which in the monopole phase, supports Abrikosov-Nielsen-Olesen (ANO) half-flux electric vortices. We showed that the ANO vortex likewise carries the 't~Hooft anomaly when the number of scalars is larger than $1$, and the vortex worldsheet either breaks charge conjugation symmetry or is gapless and saturated by the $SU(2)_1$ WZW theory. 

Finally we discussed the reduction of the $SO(3)$ symmetry by the easy-axis/plane deformations, and showed that as long as the deformation keeps the $\Z_2$ symmetry which exchanges the North and South pole of the Bloch sphere, the 't~Hooft anomaly persists. The system carries a 't Hooft anomaly between the remaining $SO(2)$ symmetry, $\Z_2$-exchange symmetry and $\Z_2$-topological symmetry. As a result the vacuum must always be nontrivial in the bulk and must break one of these symmetries.

In the case of the VBS phase, the domain wall carries the 't Hooft anomaly between the charge conjugation symmetry, the $SO(2)$ symmetry and the $\Z_2$ symmetry, which forces the domain wall theory to be nontrivial. The semiclassical regime (which is realized for large positive masses-squared of the Higgs field in the bulk) realizes the charge conjugation symmetry breaking on the domain wall. As the mass is lowered it is plausible that the domain wall undergoes the transition to the $\Z_2\times O(2)$ compact scalar model, where $\Z_2$ signifies that only even winding modes are allowed, and the system has a $\Z_2$-topological symmetry. This model can further be either in the gapless phase, or the vortex phase. 

Alternatively the bulk can be in the $\Z_2$-exchange-symmetry-broken phase. This also allows for a semi-classical description of the domain wall, which now supports the $\Z_2\times O(2)$ compact boson theory on the domain wall semi-classically. Furthermore, the vortices on the domain wall in this phase are monopoles from the bulk. Semi-classical phase is a gapless phase in this case. As the bulk is driven towards the VBS phase, the domain wall theory will likely undergo a phase transition to the vortex percolating gapped phase, which breaks the $\Z_2$-topological symmetry on the domain wall, before the bulk undergoes the phase transition in the bulk.

\section*{Acknowledgments}
We would like to thank and C. Bachas, D. Gaiotto, J. Gomis, O. Motrunich, A. Nahum, N. Seiberg, R. Thorngren,  P. Wiegmann, C. Wang, and C. Xu for useful discussions. Z.K. is supported
in part by an Israel Science Foundation center for excellence grant and by the I-CORE program
of the Planning and Budgeting Committee and the Israel Science Foundation (grant
number 1937/12). Z.K. is also supported by the ERC STG grant 335182 and by the Simons
Foundation grant 488657 (Simons Collaboration on the Non-Perturbative Bootstrap).   M.\"U. is supported by U.S. Department of Energy, 
 Office of Nuclear Physics under Award Number  DE-FG02-03ER41260.

\begin{appendix}
\section{Yang-Mills theory at $ {\theta=\pi}$ and Supersymmetric Yang-Mills} 

The Yang-Mills Lagrangian  is given by
\eqn{LagYM}{S=\int d^4x \ {\rm Tr}\left( {-1\over 4 g^2}F\wedge \star F+{i\theta\over 8\pi^2}F\wedge F\right)~,}

At $\theta=0$ or $\theta=\pi$ the model ~\eqref{LagYM} enjoys CP symmetry and $\mathbb{Z}_N$ center (1-form) symmetry. There is a mixed 't Hooft anomaly between these two symmetries at $\theta=\pi$ \cite{Gaiotto:2017yup}. 

The anomaly can be understood as follows. The CP symmetry at $\theta=\pi$ is due to the quantization of topological charge. However upon gauging the center symmetry of the $SU(N)$ theory, topological charge is no longer an integer, but can be fractional,\footnote{The simplest way to see this is to note that constant magnetic and electric fields with $1/N$ fluxes on $T^4$ are allowed in the $PSU(N)=SU(N)/\mathbb Z_N$ theory, i.e. a configuration $F_{12}=\frac{2\pi n_{12}}{N L_1L_2}T,F_{34}=\frac{2\pi n_{34}}{N L_3L_4}T$, with $T=\text{diag} (1,1,\dots,1,-(N-1))$ is a Cartan generator for which $e^{i\frac{2\pi}{N}T}\in \mathbb Z_{N}$. It follows that topological charge of such configuration is $Q=\frac{1}{8\pi^2}\int\tr{F\wedge F}=n_{12}n_{34}(N-1)/N$, which is a multiple of $1/N$ and not an integer in general. If the center is not gauged, the integers $n_{12},n_{34}$ must be multiples of $N$ which renders $Q$ an integer.} and the CP symmetry is lost. This means that there is a mixed anomaly between the CP symmetry and $\Z_N$ center symmetry \cite{Gaiotto:2017yup}. 

The anomaly can be, for instance, saturated by breaking at least one of these symmetries spontaneously (in principle it could also be that the theory is gapless or that there is long range topological order. We exclude these two possibilities from consideration even though they could be relevant for small values of $N$, especially $N=2$). The most likely scenario for sufficiently large but finite  $N$  is that the theory breaks time reversal spontaneously. This is known to be the case in the planar limit $N=\infty$, softly broken SUSY, as well as some deformations of the Yang-Mills theory \cite{Witten:1998uka,Unsal:2012zj,Anber:2013sga,Poppitz:2012nz,Poppitz:2012sw}. Therefore the zero temperature theory has two ground states and allows for a domain wall between them. 

Roughly speaking since a domain wall interpolates between the two CP vacua, the middle of the domain wall must restore the CP symmetry. Hence the anomaly inside the domain wall requires that the vacuum of the theory on the domain wall couples to the center symmetry. Therefore Wilson lines on the domain wall, which correspond to probe quarks, are deconfined. In some sense this is consistent with the general idea that the physics of the domain wall is that of the original theory in the ultraviolet\footnote{We thank N. Seiberg for emphasizing this perspective.}.

One can compute the anomaly of the theory on the domain wall precisely: it is a $\Z_N$ 't Hooft anomaly for the center symmetry, so the anomaly polynomial takes the form ${1\over N}\int_4 B^2$, where $B$ is a two-form gauge field valued in $\Z_N$. This anomaly is matched by $SU(N)_1$ pure Chern-Simons theory, where the Wilson lines are indeed deconfined and obey a simple algebra. This algebra of Wilson lines associated to probe quarks can be derived directly from the anomaly polynomial. We can compactify the domain wall theory on $\mathbb{T}^2$ and the anomaly polynomial then reduces to a mixed $\Z_N\times\Z_N$ anomaly between standard, 0-form symmetries of the quantum mechanical model. As a result, the quantum mechanical model has an $N$ dimensional ground state which is in a representation of a $\Z_N$ central extension over $\Z_N\times\Z_N$. This is just the Heisenberg group of order $N^3$. We can think about this quantum mechanical model as a particle on a torus with $N$ units of magnetic field, hence, there are $N$-fold degenerate ground  states  transforming under the Heisenberg group and the projective Hilbert space transforms under $\Z_N\times\Z_N$ as it should. (This is the degeneracy of the lowest Landau level.)   
All these properties are of course reproduced by $SU(N)_1$ Chern-Simons theory.

So far we have discussed the pure Yang-Mills theory.  If we add one adjoint Weyl fermion, the theory still has an exact center symmetry 
 $\Z_N$, as well as an anomaly free  axial $\Z_{2N}$ symmetry. The  existence of $\Z_{2N}$ symmetry is tied with the quantization of topological charge, namely, axial charge non-conservation in the $SU(N)$ theory is $\Delta Q_5= T({\rm adj})  \times  {1 \over 8\pi^2} \int 
\Tr( F\wedge F)  \in 2N \times \Z $. 
  Upon gauging $\Z_N$ center,   the topological term  is  modified into $ {1 \over 8\pi^2} \int 
\left(  \Tr F' \wedge F' -  \frac{1}{N} \Tr F' \wedge \Tr F'   \right)  \in \frac{1}{N} \Z$, where $F'$ is the field strength for $U(N)$, and can assume fractional values. Therefore, 
  $\Delta Q_5  \in 2N \times  \frac{1}{N} \Z = 2 \Z $, and one looses the axial symmetry. Only fermion number modulo two survives gauging of the center.  
  This implies that there is a mixed anomaly between these two symmetries~\cite{Dierigl:2014xta},\cite{Gaiotto:2014kfa}.  Assuming confinement (unbroken center),  the 
axial   $\Z_{2N}$ symmetry  must be broken to saturate the anomaly  and there are  $N$ vacua.  The domain walls support  Chern-Simons theories on which the quarks are deconfined.  

 There are two independent  weak coupling calculation that shows consistency with the anomaly calculation. On thermal compactification of this theory on $\R^3 \times S^1_\beta$,  where $\beta$ is inverse temperature, on small $\beta$, center-symmetry is broken, and axial symmetry is restored. 
Using similar arguments to those in~\cite{Gaiotto:2017yup},  anomaly predicts that the axial symmetry must be restored at a higher temperature than the deconfinement transition, namely\footnote{This formula also appeared in \cite{Shimizu:2017asf}, almost simultaneously with the first version of this work.}  
\begin{equation}
\beta_{\rm discrete \;  chiral}  \leq  \beta_{\rm deconfinement}
\end{equation}
  In other words, an intermediate deconfined  phase with broken chiral is  possible, but a confined phase with unbroken chiral symmetry is impossible. On circle (non-thermal) compactification on $\R^{2,1} \times S^1_L$ where fermions are endowed with periodic boundary conditions, theory preserves its center symmetry even  at  small-$L$ \cite{Unsal:2007jx, Davies:1999uw}.  In this regime, one can prove chiral symmetry breaking by semi-classical methods, and theory exhibits confinement with chiral symmetry breaking, consistent with anomaly.  There is evidence from lattice 
  simulations for the  above predictions of the anomaly, see e.g.~\cite{ Bergner:2014saa,Karsch:1998qj}.

While the discussion in this subsection is mostly a review of results that already appeared earlier, we emphasize that here we took the perspective that the anomaly polynomial itself is sufficient in order to derive deconfinement on the wall and the algebra of Wilson lines follows as well.

\end{appendix}
\bibliography{refs}

\end{document}